\documentclass[12pt,preprint]{aastex}
\usepackage{natbib,epsfig}

\begin{document}

\newcommand{\hii}{{\rm H}{\sc ii}}
\newcommand{\uchii}{{\rm UCH}{\sc ii}}
\newcommand{\iras}{{\it IRAS}}
\newcommand{\htwo}{${\rm H_2}$}
\newcommand{\nhone}{NH$_3$(1,1)}
\newcommand{\nhtwo}{NH$_3$(2,2)}
\newcommand{\nhthree}{NH$_3$(3,3)}
\newcommand{\ammonia}{NH$_3$}
\newcommand{\methanol}{CH$_3$OH}
\newcommand{\htwoo}{H$_2$O}
\newcommand{\hsixalpha}{H66$\alpha$}
\newcommand{\thcoone}{$^{13}$CO ${(J=1\rightarrow0)}$}
\newcommand{\ceioone}{C$^{18}$O ${(J=1\rightarrow0)}$}
\newcommand{\ceio}{C$^{18}$O}
\newcommand{\cseo}{C$^{17}$O}
\newcommand{\nh}{NH$_3$}
\newcommand{\um}{$\mu$m}
\newcommand{\percc}{cm$^{-3}$}
\newcommand{\persqcm}{cm$^{-2}$}
\newcommand{\emunits}{pc~cm$^{-6}$}
\newcommand{\kms}{km~s$^{-1}$}
\newcommand{\vlsr}{${\rm v}_{\rm lsr}$}
\newcommand{\msun}{M$_\odot$}
\newcommand{\mdot}{M$_\odot$yr$^{-1}$}
\newcommand{\lsun}{L$_\odot$}
\newcommand{\jyperbeam}{Jy beam$^{-1}$}
\newcommand{\mjyperbeam}{mJy beam$^{-1}$}
\newcommand{\jyperbeamkms}{Jy beam$^{-1}$km~s$^{-1}$}
\newcommand{\mjyperbeamkms}{mJy beam$^{-1}$km~s$^{-1}$}
\newcommand{\arcseconds}{$''$}
\newcommand{\pv}{P-V}

\title{The Molecular Accretion Flow in G10.6-0.4}

\author{Peter K. Sollins, Paul T. P Ho}
\affil{Harvard-Smithsonian Center for Astrophysics, 60 Garden Street, Cambridge, MA, 02138, psollins@cfa.harvard.edu}

\begin{abstract}

We have observed the ultracompact \hii\ region G10.6-0.4 with the VLA
in 23~GHz continuum and the \nhthree\ inversion line. By analyzing the
optical depth of the line as well as the kinematics, we have detected
a flattened, rotating, molecular accretion flow. We detect the fact
that the highest column density gas is more flattened, that is,
distributed more narrowly, than the lower column density gas, and that
there is some inclination of the rotation axis. The rotation is
sub-Keplerian, and the molecular gas is not in a rotationally
supported disk. We do not find a single massive (proto)star forming in
a scaled up version of low mass star formation. Instead, our
observations suggest a different mode of clustered massive star
formation, in which the accretion flow flattens but does not form an
accretion disk. Also in this mode of star formation the central object
can be a group of massive stars rather than a single massive star.

\end{abstract}


\section{Introduction} \label{sec:intro}

G10.6-0.4 is a bright, 2.5~Jy at 23~GHz \citep{ket87b}, ultra-compact
(UC) \hii\ region \citep{woo89a} at a distance of 6.0~kpc
\citep{dow80}, in an area of active star formation. The associated
\iras\ point source, \iras\ 18075-1956, has a luminosity of $9.2
\times 10^5$~\lsun\ \citep{cas86} and has colors that meet the
criteria of \citet{woo89a} for an \uchii\ region. G10.6 is known to be
embedded in a hot molecular core (HMC) \citep{bra83,ho86,plu92}. The
core is thought to contain 1200~\msun\ of gas within a radius of
0.2~pc, based on an analysis of a variety of dust continuum
measurements \citep{mue02}, and 3300~\msun\ within 1.1~pc based on
\ceio\ and \cseo\ measurements \citep{hof00}. Previous studies of the
inversion lines of \ammonia\ have determined that the molecular core
is rotating and collapsing inward toward the \uchii\ region
\citep{ho86,ket87a,ket88,ket90}. In these studies, using the \nhone\
and \nhthree\ lines, rotation is seen at size scales from 1~pc down to
0.08~pc, and infall is detected in the form of red-shifted absorption
seen against the continuum source. \methanol\ and \htwoo\ masers are
seen distributed linearly in the plane of the rotation
\citep{wal98,hof96}, while OH masers seem to lie along the axis of
rotation \citep{arg00}. In \ceioone, \citet{ho94} see $10^3$~\msun\ of
dense (n$\sim10^6$\percc), rotating gas in a flattened
(0.3$\times$0.1~pc) disk-like structure. At the highest resolution
achieved in earlier work, infall and rotation in the molecular gas
were seen simultaneously in absorption, showing that the molecular gas
was spiraling inward on size scales comparable to the size of the
\uchii\ region.

Recent observations of G10.6 hinted that it might represent a
previously unobserved mode of high mass star formation. Observations
of \hsixalpha\ from the ionized gas within the \uchii\ region indicate
that the ionized gas is also spiraling inward toward the stars at the
center of the \uchii\ region \citep{ket02a}. Subsequent theoretical
work showed that in small \hii\ regions, the gravitational effect of
the central star(s) can overcome the thermal pressure of the ionized
gas causing the molecular accretion flow to pass through the \hii\
region boundary and continue inward as an ionized accretion flow
\citep{ket02b}. In this model, the \hii\ region boundary exists as a
standing R-type ionization front within a continuous accretion
flow. These results differ from classical treatments of the pressure
driven expansion of \hii\ regions, which predict outward motion of the
ionized gas as soon as the \hii\ region is formed
\citep{str39,spitzer}. In the classical model for pressure driven
expansion, the \hii\ region boundary, after a very short phase as a
moving R-type front, will develop a characteristic double front
structure composed of an isothermal shock followed by a moving D-type
ionization front. As the \hii\ region expands, most of the displaced
molecular material remains between the shock and the ionization front
as a dense outward moving shell, which snow-plows ahead of the \hii\
region. If, however, the accretion flow passes through a standing
R-type ionization front at the \hii\ region boundary and continues
toward the star(s) as an ionized flow, as suggested by \citet{ket02b},
there will be no dense molecular layer at the boundary, and all the
molecular gas will be moving inward.

\citet{sol05a} did preliminary analysis of the data presented here,
and showed that G10.6 is accreting through its \uchii\ region. The
velocities of the molecular gas showed clear evidence of both infall
and rotation. Based on geometrical arguments, \citet{sol05a} concluded
that the infalling layer proceeded directly up to the ionization
front. They also showed a non-detection of any expanding molecular
gas. The non-detection placed such a stringent upper limit on the mass
of any expanding molecular shell that might be present, that it was
concluded that no such expansion was taking place, and that the
accretion in the ionized and molecular gas were all part of a single
accretion flow which continues across a stalled ionization front.

In this paper we present a more detailed analysis of the \ammonia\
data. We conclude that, while the accretion flow is spherical at large
radii, the rotation does cause it to flatten somewhat on the size
scale of the \uchii\ region, so that the highest column densities
appear in a thin strip. We also conclude that the axis of rotation is
inclined away from the observer in the northeast. We find that G10.6
is quite different than other young high mass stars in which disk-like
molecular structures have been observed \citep{zha98b,zha02,she99,chi04}.

\section{Observations} \label{sec:obs} \label{sec:tau}

We observed the \uchii\ region G10.6 with the NRAO Very Large Array
(VLA)\footnote{The National Radio Astronomy Observatory is a facility
of the National Science Foundation operated under cooperative
agreement by Associated Universities, Inc.} on February 1, 2002, with
the phase center at $\alpha(2000)=\rm{18^h10^m28^s.683}, \,
\delta(2000)=-19^o 55'49''.07$. We observed the (3,3) inversion line
of \ammonia\ at 23.870130 GHz with 63 spectral channels of width
48.828 kHz (0.61 \kms) for a total bandwidth of 3.125 MHz (38.7 \kms)
centered on \vlsr$=10$ \kms, and 1.3~cm continuum with a bandwidth of
15.6 MHz. The array was in the A-configuration, yielding a
uniform-weighted synthesized beam of width $0.''12 \times 0.''072$ for
a physical resolution of 0.0034$\times$0.0021~pc or 700$\times$430~AU.

We observed the quasars 3C286, 3C273 and 1733-130 for flux, bandpass
and phase calibration respectively. Self-calibration of the source
amplitudes and phases resulted in a noise level of 0.18 (0.14)
\mjyperbeam\ in the uniform (natural) weighted continuum map, and 1.9
(1.5) \mjyperbeam\ in each uniform (natural) weighted channel map,
about 3 times the thermal noise limit. The images were deconvolved by
CLEANing in the usual way with the AIPS task IMAGR. No special steps
were taken to deal with the fact that the emission area is much larger
than the synthesized beam. The total flux in the resulting natural
weighted continuum map is 2.5~Jy, which is consistent with the total
flux detected in earlier lower resolution maps of 23~GHz continuum
\citep{ket87a}. For this reason, we believe that the continuum map is
missing very little flux due to the lack of short baselines.

\begin{figure}
\epsscale{0.8}
\plotone{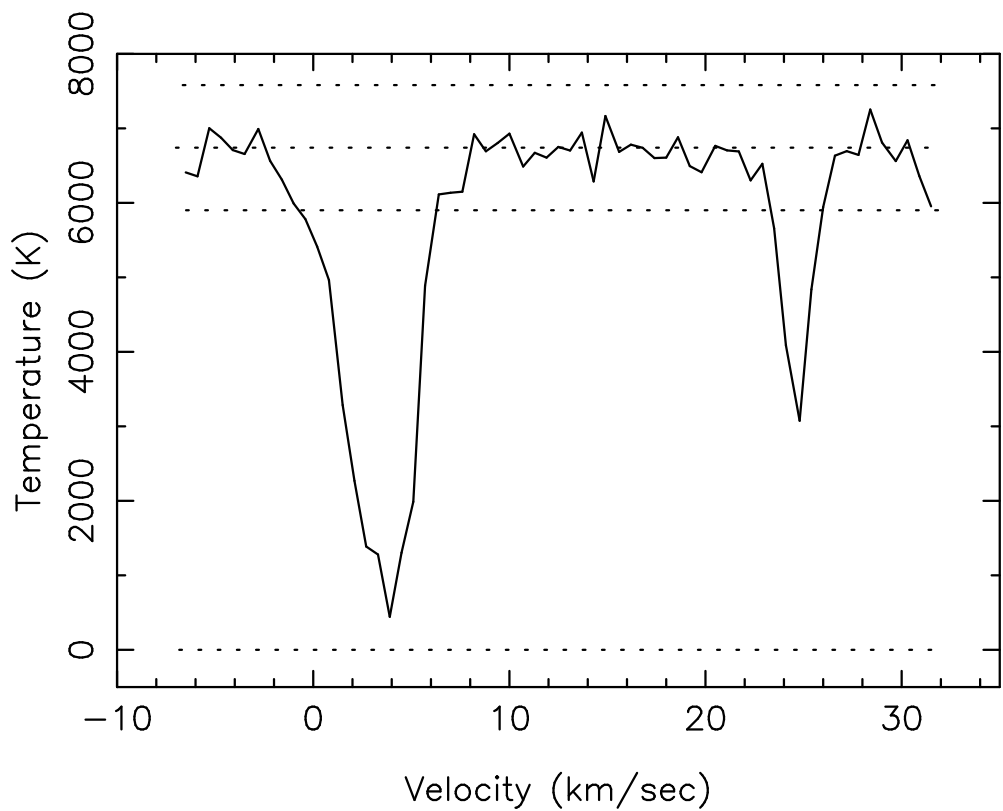}
\caption{A sample spectrum of the \nhthree\ line from near the peak of
the continuum emission. The dotted lines show the continuum level,
here 6700~K, the continuum plus and minus $3\sigma$, and zero. The
first satellite line is clearly visible, but the outer satellite is
out of the spectral window.}
\label{fig:spectrum}
\end{figure}

Expressed as a temperature, our sensitivity in a natural weighted
continuum map is about 25~K and in a natural weighted channel map is
about 280~K. The physical temperature of the molecular gas around
G10.6 is estimated to be only 110~K at the ionization front
\citep{ket90}. Thermal line emission always has a brightness
temperature less than the temperature of the gas. Thus, the brightest
possible thermal emission from the molecular gas would be
undetectable, less than $0.4\sigma$. Absorption, however, should be
detectable at a wide range of optical depths. The continuum has a peak
brightness temperature of 6900~K, and since the noise level is a
channel map is 280~K, the molecular line absorption should be
detectable at up to $25\sigma$. The quality of the self-calibration
solutions and improvements in the K-band receiver system at the VLA
have resulted in 25 times better sensitivity in our \nhthree\ channel
maps than in the previous best existing \nhthree\ data for this source
\citep{ket88}, with 3 times better spatial resolution and 2 times
better velocity resolution. A sample spectrum is shown in Figure
\ref{fig:spectrum}.

It should be noted that we have achieved the highest possible angular
resolution in studying the problem of high mass star formation. With
the VLA in its most extended array configuration, at a frequency of
23~GHz, our spatial resolution is $0''.1$. Imaging thermal (i.e.,
non-masing) molecular gas at that resolution is only possible toward
sources with strong continuum emission, and only at wavelengths which
include spectral lines useful for studying the dense gas surrounding
\uchii\ regions. Using radiation with wavelengths around a centimeter
is ideal because it is near the peak of the continuum emission from
many interesting \uchii\ regions. At lower frequency, the spatial
resolution decreases. At higher frequency the brightness temperature
of the continuum emission declines rapidly. So for optimal
backlighting from the \uchii\ region, centimeter wavelengths are
ideal. There are few thermal lines associated with high density gas in
the centimeter wavelength regime apart from the inversion lines of
ammonia which have a critical density of roughly $10^4$\percc
\citep{ho83}. In addition to its fortuitous wavelength, the hyperfine
structure (one main line, four satellite lines) of the ammonia
inversion transitions is extremely useful. Because the different
hyperfine components have well-known intrinsic line-strengths, the
ratio of the main line to a satellite line can be used to directly
calculate the optical depth (what we call hyperfine optical depth) and
column density of the ammonia in the rotational state in question, in
this case (J,K)~=~(3,3). While the optical depth of any absorption
line can be calculated directly by comparing the depth of the
absorption to the strength of the background continuum, this apparent
optical depth has limitations. If the filling factor of the absorbing
gas is less than one, the apparent optical depth decreases. Also, for
deeply embedded objects like G10.6, the main hyperfine component
easily saturates. For this reason, the satellite lines, which are much
more optically thin and do not saturate easily, are invaluable in
investigating the highest optical depths and column densities. Since
the hyperfine optical depth is calculated from the ratio of the main
line absorption to that of the satellite line, the hyperfine optical
depth is accurate even when the main line saturates, and also does not
include the effects of the filling factor. The high spatial resolution
and sensitivity to high column density gas achieved here have not yet
been possible for this sort of object in the millimeter, IR, optical,
or X-ray regimes.

\section{Results} \label{sec:results}

We report 6 key observational results. First, we find that the
absorbing molecular gas is less spatially extensive than the continuum
emission, and, on average, northeast of the average position of the
continuum emission. Second, we find that the highest column density
gas is localized in clumps, while lower column density gas is seen
over the entire face of the \uchii\ region. Third, we find that the
characteristic size scale of the infall-and-rotation kinematic pattern
of the molecular gas is larger than the characteristic size scale of
the structure in either the optical depth maps, or continuum
map. Fourth, we find that the highest optical depth gas is missing on
the southwest side of the \uchii\ region, reinforcing the idea that
the absorption is preferentially located in the northeast. Fifth, on
size scales smaller than the synthesized beam, i.e. less than $0''.1$
or 500~AU, in more than 75\% of the pixels where absorption is
detected, the filling factor of the absorbing gas is greater than
0.7. Sixth, none of the sharp edges clearly seen in the continuum
emission are seen in maps of the optical depth.

We find that the absorbing molecular gas, as located by the actual
line absorption, the apparent optical depth and the hyperfine optical
depth, is on average northeast of the continuum emission, and is
spatially narrower than the continuum emission in the direction of the
minor axis, with the hyperfine optical depth being the most skewed to
the northeast, and the most narrow. We have taken slices through the
maps shown in Figure \ref{fig:mom0panels}. The slices run
northeast-southwest, and are separated by a synthesized beam-width
($0''.14$). For each slice we have calculated the first moment to
determine the mean position of the flux in the maps, and the second
moment to determine the width of the flux in the maps. Figures
\ref{fig:meanposition} and \ref{fig:widths} show the mean positions
and widths as a function of position. The origin was chosen so that
the mean positions in the continuum map have an average of zero. In
the other three maps, the mean positions are generally northeast of
the continuum positions. For the hyperfine optical depth map, the mean
positions are on average $0''.07$ northeast of the continuum mean
positions. Only in slices which pass through the large line-width
clump, noted in Figure \ref{fig:moments}, are the mean positions of
the line absorption or apparent or hyperfine optical depth southwest
of the continuum positions. The continuum is the widest of the three
maps, while the hyperfine optical depth is the narrowest.

We find that the highest column density gas is localized in clumps,
while lower column density gas obscures the entire \uchii\
region. Comparing the two optical depth maps in Figure
\ref{fig:mom0panels}, we note that the hyperfine optical depth map
looks ``clumpier''. The flux in the map is mostly collected into peaks
about $0''.25$ to $0''.5$ (1500 to 3000~AU) in size, while the
apparent optical depth map shows more extended absorption over the
whole face of the continuum source. The large line-width clump at the
western edge is visible in both maps, as is the clump at the eastern
edge. Since the hyperfine optical depth is sensitive to much larger
optical depths than the apparent optical depth, we interpret the
difference in the maps to imply that, while there is an extended high
density envelope, the highest column densities are achieved only in
highly localized areas.

We find that the characteristic size scale of the kinematic pattern is
much larger than the characteristic size scales of fluctuations in the
optical depth or continuum maps. Figure \ref{fig:moments} shows the
first and second moments of the line absorption in the main hyperfine
component. \citet{sol05a} has interpreted the bulls-eye pattern in the
first moment map as showing simultaneous infall and rotation in a
rotating, quasi-spherical, molecular accretion flow. The velocity
field appears smooth across the absorption region, with the velocity
gradient varying slowly. The size scale of this velocity pattern is
visibly much larger than either the size of the continuum structures,
or the size of the clumpiness of the optical depth. The clumpiness in
the optical depth seems to have no effect on the velocity pattern. The
velocity pattern is established for the core as a whole, while the
optical depth appears to be picking out over-densities which do not
depart from the general flow. The map of the second moment shows that
the line width is fairly constant across most of the face of the
\uchii\ region, at around 1.8~\kms\ (FWHM = 4.2~\kms). One spot on the
western edge of the \uchii\ region shows a much larger width, $>3$\kms
(FWHM$>7$\kms). Interestingly, the bulls-eye pattern in the first
moment map shows no real effect of the anomalous large line width
clump. That location does not stand out at all in the first moment
map, which means that the broadening at this point must be symmetric
around a central velocity which fits with the overall velocity field.

The position-velocity cuts highlight both the kinematic pattern seen
in the first moment, and also the lack of high optical depth gas in
the southwest of the \uchii\ region. Figure \ref{fig:aptauposvel}
shows two position velocity cuts through the cube of apparent optical
depths. The upper panel shows a cut from northwest (negative position)
to southeast (positive position). The lower panel shows a cut from
southwest (negative position) to northeast (positive position). The
largely spherical infall can be seen in the lower panel as a backwards
``C'' shape. Only the front side of the infall can be detected since
we are seeing the line in absorption. In the upper panel, the
backwards ``C'' shape is seen again, but this time tilted, showing the
effect of rotation. The lower panel is a cut along the axis of
rotation, so only infall is seen. The upper panel is a cut in the
plane rotation, so both infall and rotation are seen. In addition, it
should be noted that the satellite line can be seen tracing all of the
absorption in the upper panel. The satellite appears everywhere in the
plane of rotation. But in the lower panel, the satellite fades out as
the cut approaches the southwest side of the \uchii\ region. The
satellite line has an intrinsic line strength of roughly 3\% of the
main line, so it traces only the highest optical depth gas. Thus the
highest optical depth gas is missing on the southwest side of the
nebula.

The filling factor of the absorbing gas is large on size scales
smaller than a synthesized beam, despite the apparent clumpiness in
the optical depth maps. We have noted above that in many pixels, the
main line saturates, i.e., in the central channels of the absorption
line, there is no detectable flux. When absorbing gas has a non-zero
filling factor, the ratio of the depth of the absorption to the
continuum strength is related to the optical depth and the filling
factor by
\begin{equation}
\frac{T_{line}}{T_{cont}} = \Phi (1 - e^{-\tau}) < \Phi
\end{equation}
where $T_{line}$ is the depth of the absorption, $T_{cont}$ is the
strength of the continuum emission, $\Phi$ is the filling factor, and
$\tau$ is the optical depth. It is impossible to tell the true optical
depth when the line has saturated, only a lower limit can be
determined. However, the above inequality shows that, no matter what
the optical depth is, saturation is only possible where the filling
factor is close to one. Furthermore, every measurement of
$\frac{T_{line}}{T_{cont}}$ puts a lower limit on $\Phi$. The lower
limit on the filling factor is 0.7 in more than 75\% of all the points
in which main line absorption is detected, and more than 0.9 in 55\%
of those points.

The sharp-edges of the emission seen in the continuum map are absent
in the optical depth maps. The ``V'' shaped cavity on the northeast
side of the \uchii\ region has very sharp edges, as does the spur to
the south. These were interpreted as being the sides of an outflow
cavity. The arcs of continuum emission to the east also have sharp
edges on the sides facing the \uchii\ region. These arcs were
interpreted by \citet{sol05a} as ionized edges of clumps of molecular
material. Photons leaking out of the central \uchii\ region could
ionize these clumps externally, naturally creating the arcs, all of
which have sharp edges pointing back toward the central source. By
contrast, none of the structures in the molecular material have such
sharp edges.

\section{Discussion} \label{sec:conclusions}\label{sec:discussion}

We draw two conclusions from the observational results. First, while
the kinematics of the accretion flow are quasi-spherical with slow
rotation, the density structure appears flattened and
disk-like. Second, the plane of the flattening is inclined to the line
of sight.

\subsection{The Molecular ``Disk''}

The densest part of the accretion flow is clearly flattened. The
flattening is clear when one compares the apparent optical depth map
to the map of the hyperfine optical depth. Figure \ref{fig:mom0panels}
shows both maps. The hyperfine optical depth, which is sensitive to
much higher column densities than the apparent optical depth, is
distributed quite narrowly along a line perpendicular to the axis of
rotation. Only the large line-width clump in the west deviates from
the mid-plane. We have calculated the width of the hyperfine optical
depth and the apparent optical depth along 15 slices parallel to the
axis of rotation (as described above). Figure \ref{fig:widths} shows
the widths perpendicular to the disk plane for each of the four maps,
continuum, velocity integrated absorption, apparent optical depth, and
hyperfine optical depth. Again, the slices which include the large
line-width clump stand out. Otherwise, the velocity integrated
apparent optical depth, which is sensitive only to lower column
density gas, is distributed more broadly, while the velocity
integrated hyperfine optical depth, which is sensitive to much higher
column densities is narrower. The average 2nd moment of the slices of
the continuum map is $0''.36$, for the apparent optical depth map is
$0''.23$, and for the hyperfine optical depth map it is $0''.19$. We
conclude that the highest density gas is collected in a flattened
structure in the mid-plane.

We make three specific predictions of what would be observed if there
were a geometrically thin, optically thick accretion disk around
G10.6, like those accretion disks seen around low-mass stars. Imagine
such a disk-\uchii\ region system schematically like the planet Saturn
and its rings, with the equator inclined relative to the line of
sight so that the south pole is visible. (We will discuss the
inclination of the rotation axis in G10.6 below) The planet is the
\uchii\ region, the disk is the rings. The northern hemisphere is
obscured by the rings, and the southern hemisphere is not. There is a
sharp edge to the obscuration, not a gradual edge, where the planet
emerges from behind the rings. In the case of a thin-thick molecular
disk around a \uchii\ region, we expect that the molecular absorption
will be strong on one side and much weaker on the other. Unlike the
case of Saturn's rings, we do not expect the obscuration to be zero in
the southern hemisphere where the disk is behind the \uchii\ region,
because the whole object is embedded in a molecular cloud. But the
difference in absorption above and below the disk should be
great. Also, if the disk is really geometrically thin compared to the
size of the \uchii\ region, we expect there to be a sharp dividing
line between the obscured side and the unobscured side, just like in
the Saturn's rings analogy. Departing from the planet-ring analogy, we
can also predict that in a disk-\uchii\ region system, the absorbing
gas in the molecular disk would be well homogenized. Such a disk would
only form if the gas were rotationally supported. So the rotation
time-scale would be much smaller than the infall time-scale, and any
inhomogeneities entering the disk would be quickly smoothed out by
differential rotation. The predictions for a thin-thick molecular disk
surrounding an \uchii\ region are a large difference in optical depths
from one side to the other, a sharp dividing line between the two
sides, and structurally smooth absorbing material.
 
G10.6 does not show evidence for a geometrically thin, optically thick
disk, and in fact it fits none of our three observational predictions
for a thin-thick disk. First, Figure \ref{fig:mom0panels} shows that
there is very high optical depth gas over most of the face of the
\uchii\ region. While the southwestern edge has less high optical
depth gas than the rest, Figure \ref{fig:mom0panels} shows no
directional preference at all for the apparent optical depth. The only
continuum emission without detectable absorption is the southern spur
(to which we will return below). Second, there is clearly no sharp
dividing line, just a general trend of the higher optical depth gas to
be thinner. Third, the absorbing material is inhomogeneous. At our
full resolution, the optical depth varies greatly on size scales
($0''.25$) much smaller than the size of the \uchii\ region ($1-2''$),
and larger than synthesized beam ($0''.1$). The existence of the
arcuate structures to the east has been interpreted as evidence that
the surrounding molecular medium is clumpy, and that the arcs are
caused by ionizing photons leaking out from the central \uchii\ region
\citep{sol05a}. We can confirm the clumpiness of the accretion flow
with the optical depth maps, which show variations in integrated
optical depth of as much as a factor of eight in a projected distance
of less than 1000~AU. While not all of the variations need to be
attributed to variations in column density, the other factors which
contribute to the optical depth, excitation temperature and ammonia
abundance, might not be expected to vary greatly in the molecular
gas. The gas distribution in G10.6 is not a geometrically thin,
optically thick accretion disk. Instead, the gas is in a flattened,
slowly rotating, molecular accretion flow.

Compared to the central mass, the mass of the accretion flow is
appreciable. Based on the hyperfine optical depth, and assuming a
temperature and ammonia abundance, we can calculate the total
molecular mass seen in absorption in the accretion flow. Using 150~K
for the excitation temperature of the gas \citep{sol05a}, the peak
column density of ammonia is greater than $1.2 \times
10^{17}$~\persqcm. This a lower limit because there are points at
which the absorption in the satellite line saturates. Our integral
over the entire map therefore gives a lower limit. Assuming the
ammonia abundance is $10^{-7}$ relative to \htwo\ \citep{van98}, and
adding a factor of 2 since we only detect the front half of the
accretion flow in absorption, the total molecular gas mass in the
molecular accretion flow is greater than 72~\msun. The assumed
abundance is the largest source of error here and could be wrong by as
much as an order of magnitude in either direction. Using the radius of
the \uchii\ region to set the size scale, the implied mass accretion
rate is 0.02~\mdot. \citet{sol05a} calculate that the central mass
responsible for the infall is roughly 150~\msun. It is entirely
possible that the mass of the accretion flow is comparable to the
central stellar mass.

The total continuum flux is 2.44~Jy, so assuming constant density, and
electron temperature of 10,000~K and a physical size of 8500~AU, the
mass of the ionized gas is 0.22~\msun. Just by looking at the
continuum map it is clear that the density is not uniform, and since
we know there is ongoing accretion, the density profile should be
proportional to $r^{-3/2}$. The mass, however, depends strongly on the
total size of the region in question. So the more spread out ionized
gas will dominate the mass. For example, even if we associate all the
emission from the marginally resolved peak of the continuum emission
with a single density enhancement, the most ionized gas mass we can
possibly associate with the peak is 0.0035~\msun. \citet{ket03}
pointed out that, when estimating the Lyman continuum flux necessary
to achieve ionization balance, the density gradient can be very
important. This is because the recombination rate is proportional to
density squared. Because mass is proportional to density only to the
first power, the total mass is less sensitive to small high density
pockets, and will be dominated by the larger scale structures. Only
for a density profile steeper than $r^{-2}$ will the mass be dominated
by smaller radii rather than larger radii.

\subsection{Inclination of the ``Disk''}

\citet{sol05a} determined that the rotation axis of the molecular
accretion flow points northeast when projected into the plane of the
sky. Based on our data, the axis of rotation appears to be tipped away
from the observer in the northeast. Even though the molecular gas is
not in a rotationally supported, geometrically thin disk, the gas
distribution is flattened, with denser gas collected in the plane of
the equator of the system. Since that plane is tipped, we expect the
densest gas to be preferentially in the northeast. We have seen a hint
of this already in Figure \ref{fig:aptauposvel}, where the absorption
from the main hyperfine component extends right down to the southwest
edge of the continuum, while the absorption from satellite component
does not. To test for this inclination quantitatively, we have
analyzed the maps in Figure \ref{fig:mom0panels}. For each of the four
maps we have made fifteen slices parallel to the axis of rotation,
southwest-northeast. Then along each slice we calculate a
flux-weighted average position, i.e., the first moment of the
slice. In the continuum map, the average positions follow the line
closely, and do not systematically deviate in one direction or the
other. By contrast, in the hyperfine optical depth map, the average
positions are all to the northeast, except at the two points where the
anomalous large line-width clump has dragged the average to the
southwest. The highest density gas is, on average, $0''.08$ northeast
of the projected mid-plane of the continuum, and farther to the
northeast if the large line-width clump is excluded. Because the area
over which the optical depth can be calculated is defined by the
extent of the continuum emission, the mean positions cannot be wildly
different. This emphasizes the significance of the offsets in position
of the hyperfine optical depth from the continuum. These offsets are
direct evidence that the ``disk'' is tilted, and the axis of rotation
is inclined. Using the radius of the \uchii\ region, $1''.1$, as a
lower limit for the radius of the disk, a $0''.08$ offset is
consistent with a tilt of the disk of $4^\circ$. Other direct evidence
for this inclination has been found by \citet{ket05}, who have
detected red-shifted \hsixalpha\ emission to the northeast of the
\uchii\ region, within the notch in the continuum emission on that
side.  They have interpreted that gas as an outflow.

Another clue as to the inclination of the rotation axis is the
strength of the absorption on the narrow spurs of continuum. On the
northeast side of the continuum source there is the ``V''-shaped notch
mentioned above. \citet{sol05a} interpret this as a possible outflow
cavity, and \citet{ket05} have confirmed this. On the southwest side,
\citet{ket05} detect no corresponding blue-shifted outflow, but the
continuum does show a sharp edged spur, reminiscent of the notch in
the northeast. While the spur-like structures in the northeast show
strong absorption in both apparent and hyperfine optical depth, the
spur in the southwest shows only weak absorption in the apparent
optical depth map, and no detectable hyperfine optical depth. This is
further evidence that the high density gas in front of the continuum
source is in the northeast because of a flattened density profile and
the rotation axis being tipped away from the observer in the
northeast. However we should note a possible alternate explanation. We
cannot rule out the possibility that the southern spur is the
limb-brightened edge of a photo-ionized molecular clump, just like the
arcs to the east. In that case, the clump might be closer along the
line of sight, not physically associated with the main \uchii\ region,
and therefore not obscured by the densest molecular gas.

\subsection{A New Phase of Massive Star Formation}

The data on G10.6 are unique in the study of accretion onto massive
stars because of their spatial resolution, and because the
interpretation is not model dependent. Disk-like structures have been
detected around a number of very early B type stars.  \citet{chi04}
detected a 10000~AU, morphologically disk-like structure at 550~AU
resolution. Kinematic observations at 13000~AU resolution show that
the disk is rotating on that larger size scale. In \iras\ 20126+4104,
\iras\ 18089-1732, AFGL5142, flattening and rotation in dense
molecular gas has been detected at roughly 5000~AU resolution. In all
three of those cases the sources have infrared luminosities
corresponding to early B stars \citep{zha98b,zha02,beu04a}. In
G192.16-3.82, \citet{she99} find a velocity pattern consistent with
rotation in water maser spots at a 1000~AU spatial scale, also around
an early B star. In all these cases, molecular gas is found to be in
rotation, in some cases apparently Keplerian rotation, around early B
type or even late O type stars. All are consistent with the existence
of rotating, molecular accretion disks which, in many respects, are
larger versions of the disks observed around low-mass protostars. In
contrast to previous work, in G10.6 we have 500~AU physical resolution
in the thermally emitting molecular gas. The spatial resolution is
enough to completely resolve the motions involved. The kinematics and
optical depths are fairly unambiguous. The densest gas is flattened,
and the velocities clearly show infall and slow rotation.

G10.6 itself contrasts with objects from previous massive-star-disk
studies in that it cannot be interpreted as a scaled up version of
low-mass star formation. All the cases cited above (\iras\ 20126, \iras\
18089, G192.16, M17, and AFGL5142) are consistent with the central
object being a single stellar system of up to 20~\msun. In most of
these objects the existence of a bipolar outflow indicates ongoing
accretion. The analogy to the formation mechanisms of low-mass stars
is fairly straightforward, scaled up in size and mass, although a key
difference is ratio of disk mass to stellar mass, small for low mass
stars, but apparently large for high mass stars. In G10.6, the central
source is at least 150~\msun, close to $10^6$~\lsun, and is almost
certainly not a single star or binary. At the edge of the \uchii\
region, at a radius of 5000~AU, the molecular gas is clearly moving
inward, and \citet{ket02a} detect inward motions in the ionized gas
down to radii of less than 1000~AU. While we cannot say how or if the
inward moving gas actually accretes onto one or more of the central
stars, we can say with great certainty that inward motion continues in
the molecular gas from the 0.5~pc scale \citep{ho86,ket88} down to
thousands of AU. This is a single continuous accretion flow traceable
over two orders of magnitude in size, toward a group of young massive
stars. This suggests a completely different phase or mode of star
formation than that seen in low mass stars, or in the preceding
examples of individual massive young stars.

\section{Summary}

We have utilized the strong 23~GHz radio continuum emission from the
\uchii\ region G10.6-0.4 to serve as a backlight for examining the
foreground molecular material seen in absorption. Using the VLA, we
have achieved very high angular ($0''.1$) and spatial (500~AU)
resolution.  In the past, such resolutions have not been possible for
studying the circumstellar environment of massive young stars.  Making
use of the hyperfine structure of the \nhthree\ inversion line, we are
sensitive to optical depths of up to 80. This allowed us to
investigate the structure of the densest circumstellar material. We
find that in the densest material, the structure is flattened, with an
aspect ratio of 5. The structure is displaced with respect to the mean
continuum emission, consistent with a tilt of the disk along the line
of sight at $4^\circ$, away from the observer in the northeast. The
flattened structure has a mass of 72~\msun, much larger than the
ionized gas in the HII region of 0.2~\msun. The velocity pattern
within the circumstellar material, as well as its clumpiness, suggest
a dynamically collapsing structure which is not centrifugally
supported. The implied infall rate is very high, on the order of
0.02~\mdot. The kinematics of the circumstellar material, which agree
with the kinematics of the ionized gas within the \hii\ region,
suggest that this infalling material continues across the ionization
front. Because we do not know how much mass is actually being accreted
by the stars, or how much mass is leaving the system in the outflow,
it is impossible to know for sure whether G10.6 is in quasi-static
equilibrium, or if it is evolving dynamically. However, the very high
mass and luminosity involved mean that this is a different type of
object than the individual high-mass protostars which have been
investigated in the past.

The authors would like to acknowledge Eric Keto and Qizhou Zhang for
their helpful comments and suggestions in the preparation of this
paper.

\bibliographystyle{apj}
\bibliography{bib_entries}

\begin{figure}
\caption{The continuum map and the maps of the zeroth moments of the
line absorption, the apparent optical depth of the main hyperfine
component, and the hyperfine optical depth. In all four frames, the
contour is the 1~\mjyperbeam\ continuum level. The color-scales are
all linear, running from -1 to 45~\mjyperbeam\ for the continuum, from
5 to -200~\mjyperbeamkms\ for the line absorption, from 0 to 14~\kms\
for the apparent optical depth, and from 0 to 320~\kms\ for the
hyperfine optical depth. Notice that the apparent optical depth is
seen over the entire face of the continuum source, while the hyperfine
optical depth is concentrated in a thin strip in the plane of
rotation. The units of the optical depth maps here are \kms\ because
optical depth itself is unitless, while the maps are of optical depth
integrated over velocity.}
\end{figure}

\newpage
\addtocounter{figure}{-1}

\begin{figure}
\epsscale{1.0}
\plotone{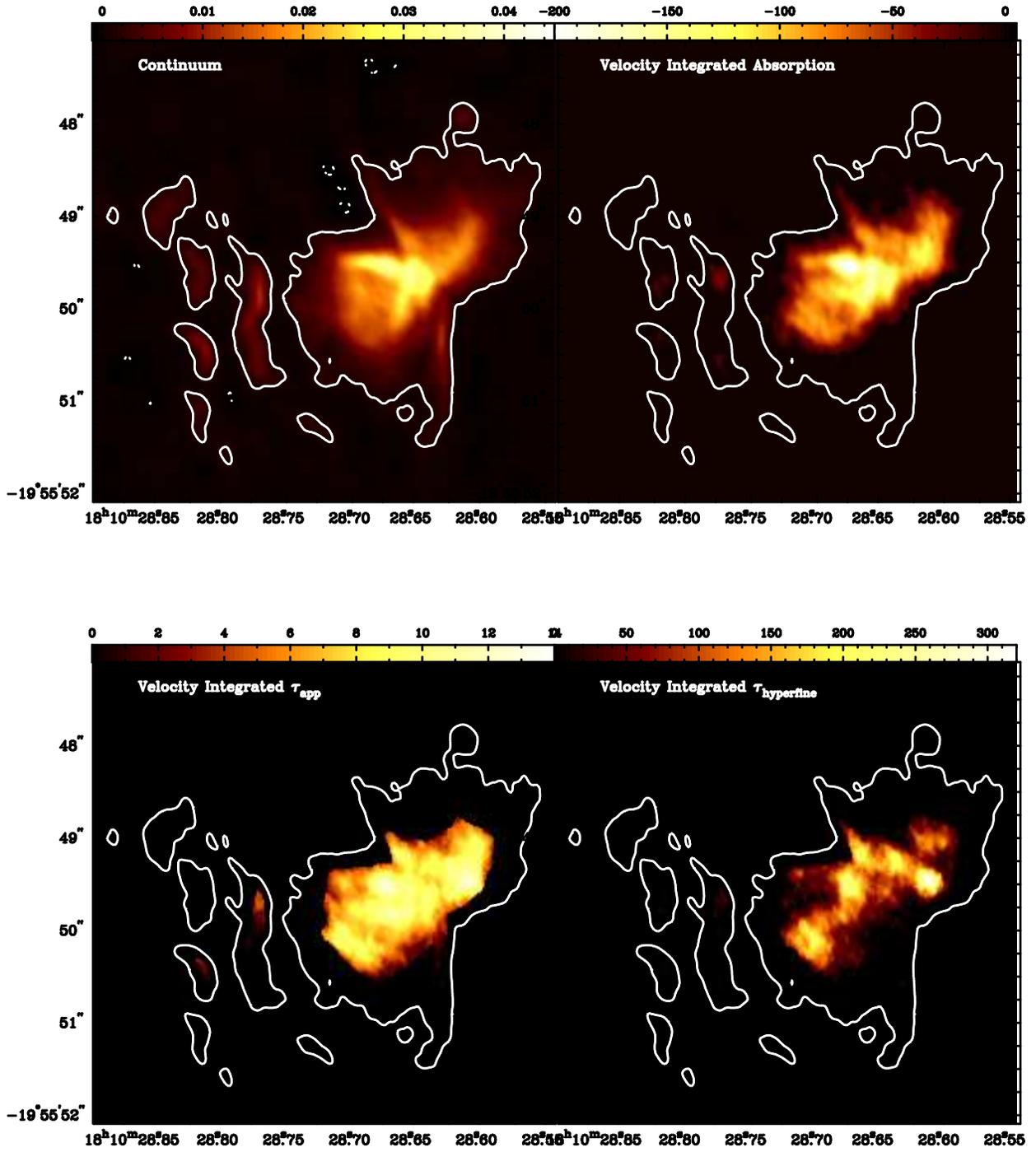}
\label{fig:mom0panels}
\caption{Continued}
\end{figure}

\begin{figure}
\epsscale{0.7}
\plotone{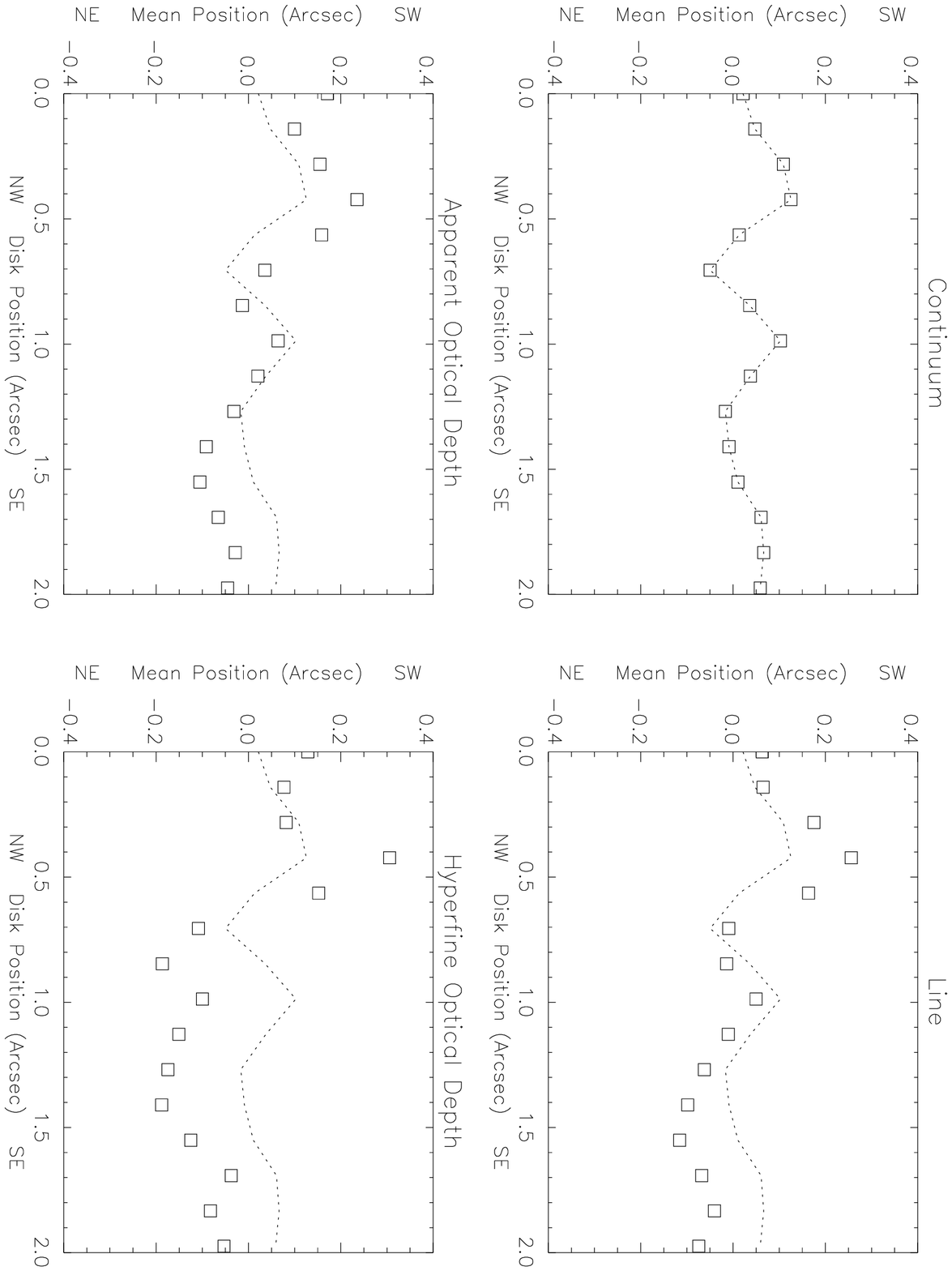}
\caption{Plots of the mean position of the flux in the continuum map
and the velocity integrated line, apparent optical depth, and
hyperfine optical depth maps relative to the mid-plane of the \uchii\
region. The squares are the mean positions, while the dotted line
shows the continuum mean positions for reference. In integrating over
velocity only the main hyperfine component is included. Except for the
position of the large line-width clump, the hyperfine optical depth
mean positions are northeast of the mid-plane. The apparent optical
depth and the raw absorption line have mean positions closer to the
continuum mean positions.}
\label{fig:meanposition}
\end{figure}

\begin{figure}
\epsscale{0.7}
\plotone{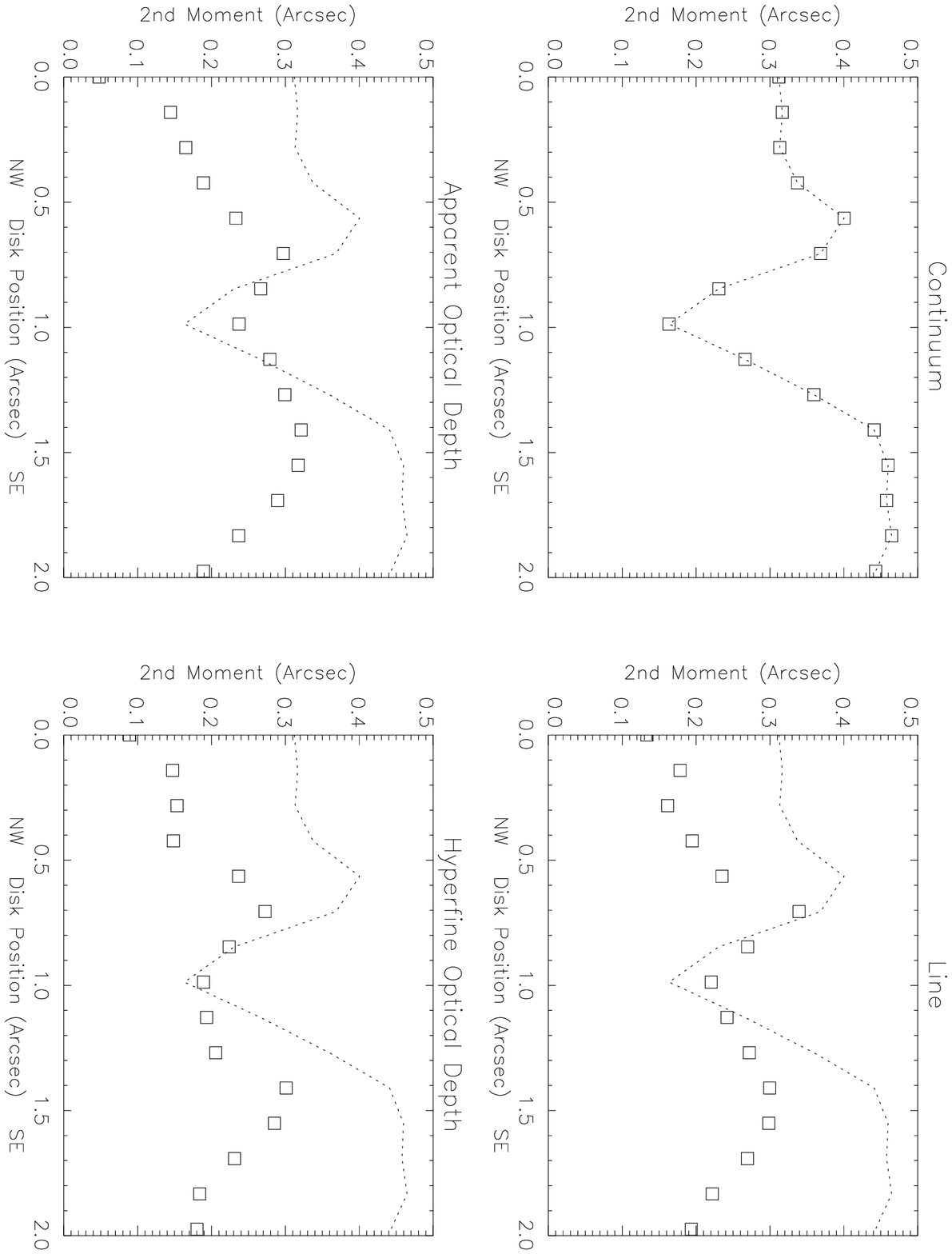}
\caption{Plots of the second moment of the flux in slices of the
continuum map and the velocity integrated line, apparent optical
depth, and hyperfine optical depth maps relative to the mid-plane of
the \uchii\ region. The squares are the widths while the dotted line
is widths of the continuum, plotted for reference. In integrating over
velocity only the main hyperfine component is included. The emission
is always narrowest near the outflow cavities where the continuum is
``pinched''. The hyperfine optical depth is the narrowest.}
\label{fig:widths}
\end{figure}

\begin{figure}
\epsscale{1.0}
\plotone{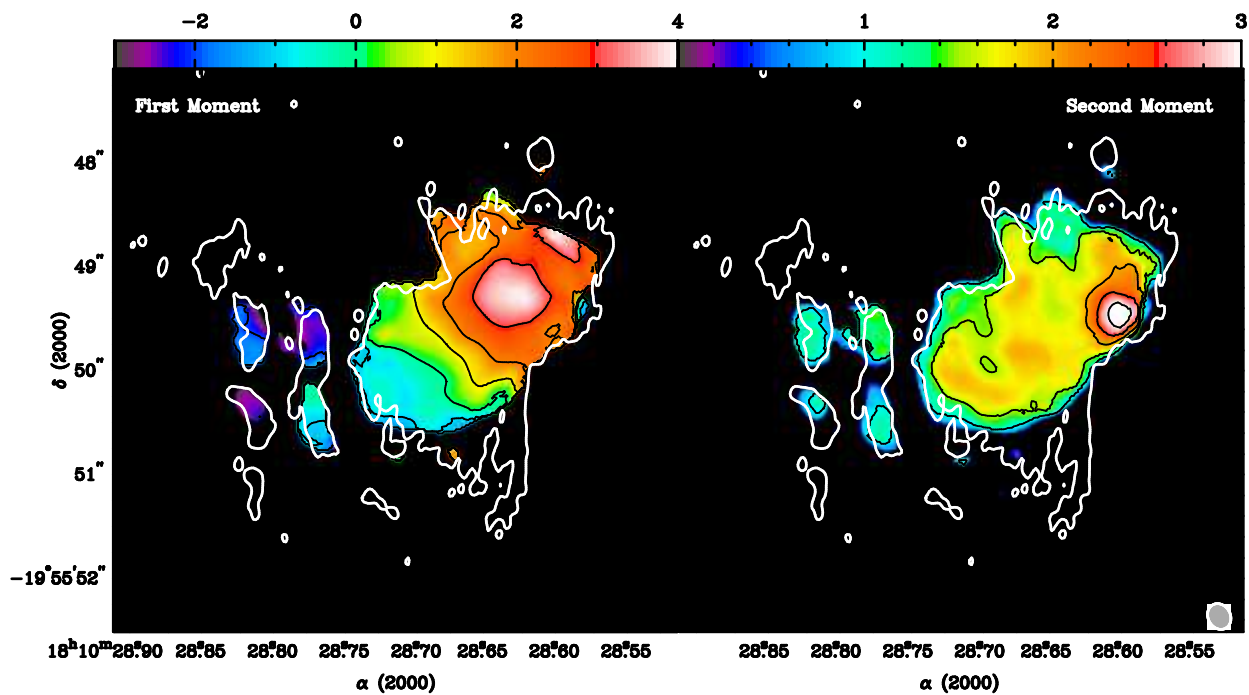}
\caption{The map of the first (left) and second (right) moments of the
main hyperfine component of the \nhthree\ line with one contour from
the continuum map. For the first moment, the colors range from \vlsr =
-3 to +4 \kms. For the second moment, the colors range from 0 to
3~\kms. For a pure Gaussian line, the full width of half-maximum is
2.35 times the second moment. The contour, shown for reference, is the
0.8 \mjyperbeam\ contour from the 1.3~cm continuum map. The line data
were mapped by the AIPS task IMAGR with natural weighting and a u-v
taper at 750~kilolambda so that the resulting synthesized beam
($0.''26 \times 0.''22$) better matched the size scale of velocity
pattern.}
\label{fig:moments}
\end{figure}

\begin{figure}
\epsscale{1.0}
\plotone{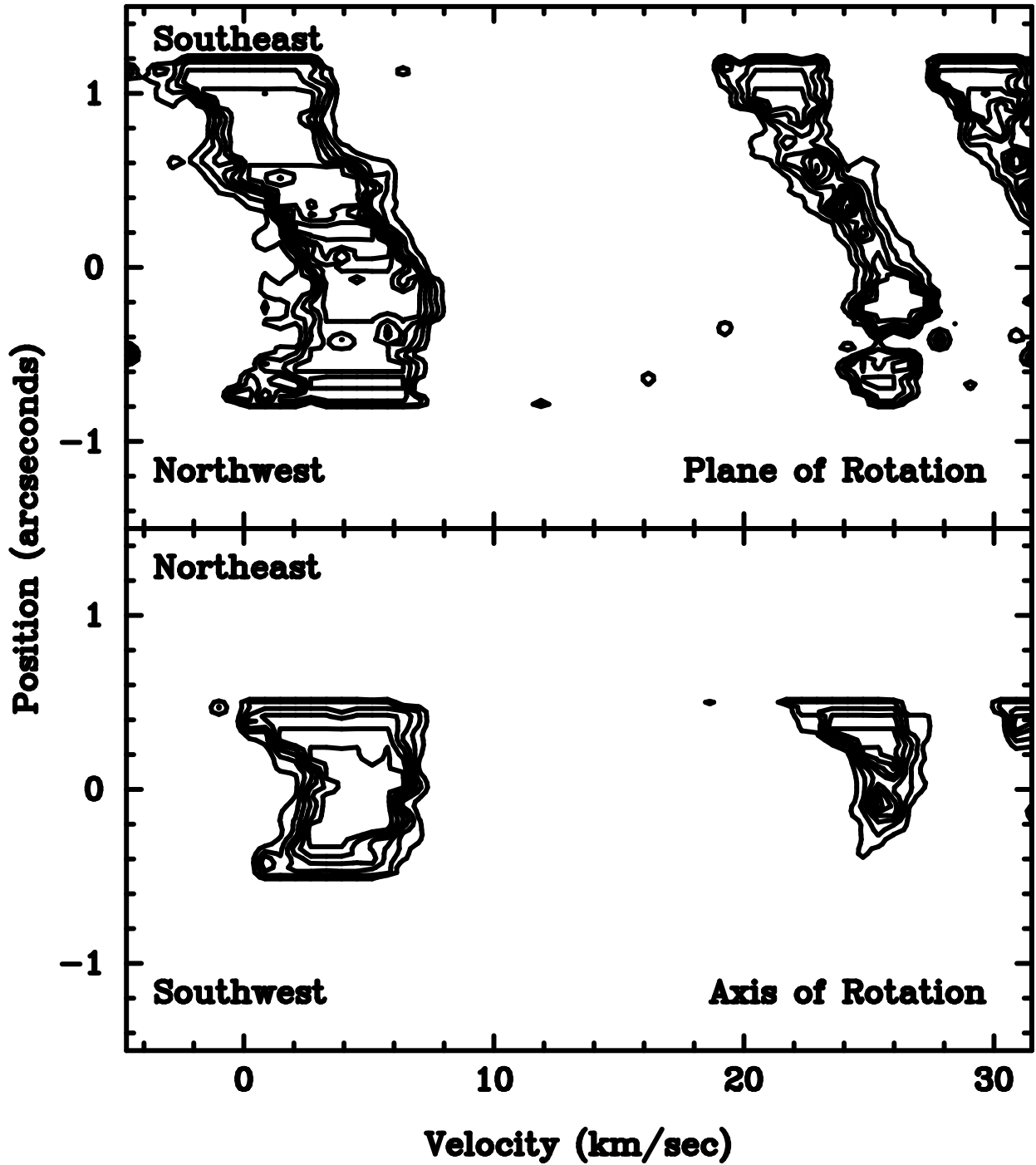}
\caption{Position velocity diagrams of two cuts through the cube of
apparent optical depths of the \nhthree\ line. The contours are at
$\tau = 1, 2, 3, 4, 5, 6, 7, 8, 9, 10 \times 0.3$. Notice that in the
plane of rotation, the satellite has roughly constant optical depth at
all positions. Along the axis of rotation however, the satellite line
fades out in the southwest indicating the presence of higher density
gas in the northeast. This is due to the inclination of the rotation
axis to the line of sight.}
\label{fig:aptauposvel}
\end{figure}

\end{document}